# Title: Tunable plasmons in ultrathin metal films

Rinu Abraham Maniyara[1]†, Daniel Rodrigo[1]†, Renwen Yu[1], Josep Canet-Ferrer[1], Dhriti Sundar Ghosh[1], Ruchirej Yongsunthon[2], David E. Baker[2], Aram Rezikyan[2], F. Javier García de Abajo[1,3]*, Valerio Pruneri[1,3]*

**Affiliations:**

[1] ICFO- Institut de Ciències Fotòniques, The Barcelona Institute of Science and Technology, 08860 Castelldefels (Barcelona), Spain

[2] Corning Research and Development Corporation, Sullivan Park, Corning, New York 14831, United States.

[3] ICREA- Institució Catalana de Recerca i Estudis Avançats, 08010, Barcelona, Spain.

* Correspondence to: valerio.pruneri@icfo.eu, javier.garciadeabajo@nanophotonics.es

† Equal contribution

**Abstract:**

The physics of electrons, photons, and their plasmonic interactions changes greatly when one or more dimensions are reduced down to the nanometer scale[1]. For example, graphene shows unique electrical, optical, and plasmonic properties, which are tunable through gating or chemical doping[2–5]. Similarly, ultrathin metal films (UTMFs) down to atomic thickness can possess new quantum optical effects[6,7], peculiar dielectric properties[8], and predicted strong plasmons[9,10]. However, truly two-dimensional plasmonics in metals has so far elusive because of the difficulty in producing large areas of sufficiently thin continuous films. Thanks to a deposition technique that allows percolation even at 1 nm thickness, we demonstrate plasmons in few-nanometer gold UTMFs, with clear evidence of new dispersion regimes and large electrical tunability. Resonance peaks at 1.5-5 μm wavelengths are shifted by hundreds of nanometers and amplitude-modulated by tens of per cent through gating using relatively low voltages. The results suggest ways to use metals in plasmonic applications, such as electro-optic modulation, bio-sensing, and smart windows.

**Main text:**

Since ancient times, plasmons in nanoparticles of noble metals such as silver and gold have been used to color glass, culminating during the last two decades with a remarkable broadening of the use of plasmon excitations triggered by an improved understanding of their origin and behaviour, as well as by the availability of more sophisticated means to synthesize and pattern the metals[11–13]. New applications promise to have an impact on the optical industry: for example, super lenses allowing unprecedented sub-diffraction-limited optical imaging[14], metasurfaces providing on-chip functionality in ultrathin form factor[15], light modulation[16], compact biosensors[17] and electrochemical effects that can be used in smart windows[18]. All these plasmonic applications would unquestionably benefit from active electro-optic tunability. This is the case of graphene, a two-dimensional material, where the surface carrier density ($n_S$) regulating the plasmonic response can be changed by applying an external electric field (voltage)[2,3]. In metals, the bulk carrier density ($n_B$) is large and in order to be able to change $n_S$ significantly one needs to achieve very small thickness ($t$) values ($n_s=n_B t$). The difficulty of making continuous films with sufficiently small $t$ over a large area has so far prevented the demonstration of electro-optic tunable plasmons in metals. Here we show that, similarly to graphene[19], UTMFs with a sufficiently low nanometric thickness can support two-dimensional plasmons. In particular, we achieve dispersion and large optical tunability by electrical gating. These experiments are made possible by a new deposition technique in which copper is used as a seed layer to produce large areas of gold UTMFs. The technique (physical vapor deposition) avoids the problem of island-like growth of unseeded gold at small thickness, giving rise to percolated films, and has the crucial advantage of being industrially scalable.

Direct evaporation of gold on substrates such as glass and other inorganic materials (for example, $CaF_2$) causes the gold UTMF to grow in metallic islands of irregular shape (Volmer-Weber growth mode) during the initial growth stages due the poor wetting of gold[20,21] (Fig. 1a). The key idea introduced in our UTMFs fabrication is the use of a sputtered copper seed layer with thickness ~1 nm, which has already been shown to produce lower percolation threshold in silver films[22]. In this work, the copper seed layer is exposed to air and is likely to undergo oxidation before the gold evaporation. The resulting percolated and polycrystalline gold UTMF is shown in Fig. 1b. For a nominal thickness $t=3$ nm determined by extrapolation of the deposition rate, in turn calibrated by a quartz crystal microbalance, we find a roughness $R_q= 0.22$ nm measured with atomic-force microscopy (AFM) at 3 nm lateral resolution (Fig. 1b). Long-range continuity of the seeded gold film is observed in scanning electron microscope (SEM) images at 3 nm resolution, in contrast to the disconnected geometry observed in unseeded films. We measured an average geometrical thickness of 4.42 nm using AFM for deposition of a mass-equivalent thickness $t=3$ nm of gold seeded by 1 nm of copper on a 285nm thick native silica oxide on silicon substrate. The AFM thickness value is also consistent with scanning transmission electron microscopy (STEM) images with sub-angstrom resolution, which additionally confirm good continuity of the film. The fact that the geometrical thickness is slightly larger than the total mass-equivalent thickness (4nm) could be due to the formation of copper oxide with a volume larger than the original metal.

To characterize the long-range connectivity and the percolation thickness of gold UTMFs, we study visible and near-infrared transmission spectra for different $t$ (Fig. 1c and d), as well as electrical properties (Fig. 1e). The minimum nominal (mass-equivalent) gold thickness for which a gold UTMF becomes physically connected and electrically conductive (percolation thickness) is $t=1$ nm and $t=7$ nm for seeded and unseeded growth, respectively. For $t=1$ nm, the seeded gold UTMF has a sheet resistance of ~1.5 k$\Omega/\square$, which is comparable to single layer graphene, while for $t=3$ nm it substantially decreases to 74 $\Omega/\square$. We also observe that for $t \geq 3$ nm, the electrical scattering time $\tau$ of the seeded gold UTMFs is in all cases several times smaller than in bulk metal.

The unseeded gold UTMFs present a transmission dip at approximately $\lambda=600-650$ nm for low coverage (Fig. 1 c), which we attribute to localized optical modes as a signature of the presence of electrically isolated metallic islands[23,24]. For $t \geq 7$ nm (the percolation thickness), the noted resonance starts disappearing and the transmission at longer infrared wavelengths becomes progressively lower. The measured spectra for the Cu-seeded gold UTMFs (Fig. 1d) become noticeably different from the unseeded films at low coverage, in which the transmission decreases at longer wavelengths, exhibiting a typical behaviour of continuous metals down to $t$=1 nm, and therefore confirming an unprecedentedly low percolation depth. In general, above percolation threshold, previous studies[8,25] show that the long-wavelength optical response of ultrathin gold films can be described using the Drude model with adjusted inelastic scattering rates[10,26]. As a result of quantum confinement in the vertical direction (along the film normal), previous theoretical work showed that single crystalline ultrathin gold films display strong anisotropy between the in-plane and out-of-plane permittivities, which might lead to a significant modification of the surface plasmon dispersion[27,28]. However, in agreement also with previous experimental studies[8,29], such anisotropy does not play a significant role in our experiments due to the fact that the samples are polycrystalline.

Next we study the infrared plasmonic properties of seeded gold UTMFs with $t$ down to 3 nm. To this end, we pattern them into nanoribbon structures of width $W$ and period $P$, which enable an efficient excitation of a localized surface plasmon resonance (LSPR), and specifically a first-order dipolar mode[19], by illumination with normally-incident infrared light (Fig. 2a). The nanostructuring of gold UTMFs on infrared-transparent $CaF_2$ substrates is carried out with electron beam lithography and argon reactive ion etching (Fig. 2b). We have fabricated two chips with two replicas per chip for a nominal Au thickness of 3 nm. The resulting transmittance and reflectance for the sharper set of spectra are shown in Fig. 2c for $t$=3 nm, $W$ from 200 nm to 800 nm and $P$=1.5$W$, exhibiting strong resonances at near- and mid-infrared wavelengths. Such resonance peaks reveal that, despite the ultrathin nature of the film, the level of optical damping is low to support plasmonic modes. We present electromagnetic simulations in Fig. 2c (dashed curves) for comparison, obtained with a Drude damping $\Gamma$=0.19 eV for 3 nm gold films, although quantitatively similar results are obtained for higher values of $\Gamma$, so we cannot determine an accurate value for this parameter, which is several times the one for bulk metal. We remark that the plasmonic dispersion of thin films are rather independent of their detailed nanoscale morphology in the current regime of deep-subwavelength thickness (that is, porous films with a geometrical thickness up to a factor of 2 higher than the mass-equivalent thickness $t$ would result in the same plasmon dispersion within the accuracy of the present study).

To further understand the nature of these plasmonic modes, we have measured the spectral response of nano-patterned seeded gold UTMFs with mass-equivalent thicknesses $t$=6 nm and 15 nm (see selected spectra in Fig. 2d). We observe that for these larger thicknesses the plasmonic resonance shifts toward shorter wavelengths. This is a well-known plasmonic effect, typical of metals in the visible wavelength range[10]. The behaviour becomes clearer by mapping the ribbons resonance obtained from experiments into a dispersion diagram (Fig. 2e), where each point is obtained from the experimental resonance frequency of each UTMF nanoribbon array given the wavevector condition $k$=$\pi$/$W$[30]. For large thickness ($t$=15 nm) the ribbon dispersion follows closely the light line dispersion (accounting for the surrounding refractive index). However, as the metal layer becomes thinner ($t$=3 nm) there is a clear dispersion bending and a deviation from the light line, which is also predicted by the theoretical loss function that evaluates plasmonic energy dissipation (Fig. 2f). The bending observed in the dispersion diagram reveals the plasmonic nature of the infrared resonance modes in UTMFs when the thickness becomes sufficiently small[30]. Importantly, the dispersion bending takes place at metal thicknesses for which the UTMF becomes infrared-transparent (see inset in Fig.2e), indicating that the plasmonic effects arise when the UTMF is thinner than the penetration depth of the infrared field. The physical origin of the measured plasmonic response lies in the low surface carrier density of UTMF when the thickness $t$ becomes sufficiently small. The discrepancy in the dispersion diagram shown in Fig. 2f originates in the fact that the theoretical calculation is performed for the propagating surface plasmon modes supported in the ultrathin gold film[26], which deviate from the experimental results for the localized surface plasmon modes supported in individual ribbons. After renormalization of the ribbon width, we achieve improved agreement between experiment and theory.

Next, we explore the electrical gating of plasmons in UTMFs. Considering the three-dimensional bulk carrier (free electron) density of gold $n_B \sim 5.9 \times 10^{22}$ cm$^{-3}$, its Fermi energy ~ 5.53 eV translates into a Fermi wavelength around 0.52 nm, which is still much smaller than the minimum film thickness ($t \sim 3$ nm) studied for plasmonic response in this work. Generally, in order to realize a two-dimensional electron gas, one needs to maintain the condition $n_B \times d^3 \ll 1$, where $d$ is the vertical dimension of the system (here, the film thickness). Given the large density of electrons in gold, such condition is not satisfied in our experiments, and one can use an effective surface carrier density $n_S = n_B \times t$, which has a value of ~ $1.8 \times 10^{16}$ cm$^{-2}$ for $t=3$ nm. As illustrated in Fig. 3a, the seeded gold UTMF ribbons are spin coated with an ion gel composed of EMIM-TFSI and PS-PEO-PS polymer (see Methods) and an electrostatic voltage ($V_g$) is applied between the ribbons and an external top electrode. We show in Fig. 3b (left, experiments; right phenomenological theory) infrared spectra of selected ribbon arrays with mass-equivalent thickness $t$ of 3 nm, 6 nm, and 15 nm for different gate voltages from $V_g = +2$ V to $-2$ V. We observe in the experiments that the plasmonic resonance remains constant for $t=15$ nm, it can be slightly modulated for $t=6$ nm, and it is highly tunable for $t=3$ nm. The corresponding experimental wavelength shift is plotted in Fig. 3C and reaches $\Delta\lambda=200$ nm for $t=3$ nm and a smaller value of $\Delta\lambda=50$ nm for 6 nm, whereas for $t=15$ nm it is below the precision of experiment. The amplitude and sign of these changes depend on those of the applied voltage and, in some cases, a hysteresis occurs. Similar hysteresis was observed in electrical resistance measurements of gold thin films using ion gel and low gate voltages as in our case[31], which the authors attributed to oxidation/reduction of gold. This can be attributed to the inherent nature of the ionic gel (i.e., a slow polarization response time of the ions, and charge trapping at the interface with the ion-gel[32]). The wavelength shift is equivalent to a substantial change in the surface carrier density as a result of voltage gating. The effective surface carrier density change used as a fitting parameter of the experimental results in Fig. 3, does not necessarily corresponds to a change of bulk carrier density (number of electrons per unit volume) induced in the metal. For negative gating voltages there is a decrease in effective surface density of conduction electrons, which in turn produces a redshift of the plasmonic resonance. From our infrared simulations (Fig. 3d) we observe that the wavelength shifts can be explained by a maximum effective surface carrier density variation up to $3 \times 10^{15}$ carriers/cm$^2$ (that is, from $V_g=+2$ to -2V), which is consistent with the modulation observed in electrical transport experiments on thicker (tens of nm) gold films by independent groups [31,33]. For this maximum surface carrier density modulation, the relative variation of effective surface carrier density in our experiments is 17%, 8%, and 3% in films of $t=3$ nm, 6 nm, and 15 nm, respectively, which explains the widening of the wavelength tuning range for lower UTMF thicknesses. These modulations occur over a characteristic response time about 1 min after application of the gating potential.

In addition to the wavelength shift, there is also a very significant modulation of the optical transmission, especially for $t=3$ nm, in which the resonance peak amplitude can be tuned from 39% to 55% transmittance. This large modulation cannot be explained by a surface carrier density change alone. Indeed, simulations indicate that for $t=3$ nm, the variation in effective surface carrier density variation is responsible for less than one-quarter of the transmission modulation range. Instead, the additional mechanism dominating amplitude modulation could be associated with an effective modification of optical damping, affecting the amplitude and width of the plasmonic peak. Our simulations show qualitative agreement with experiment when the Drude damping $\Gamma$ for $t$=3 nm changes from 0.22 eV to 0.36 eV when varying the voltage $V_g$ from +2 V to -2.0 V, while for $t$=6 nm it needs to be changed from 0.13 eV to 0.18 eV. The fact that the largest damping modulation takes places for the thinnest UTMFs indicates that surface scattering may be the dominant damping channel at such small thicknesses.

We explore also the tunability of UTMFs at different wavelengths by adjusting the ribbon width $W$. In Fig. 3e we show the transmission amplitude and resonance wavelength of the plasmonic peak for varying voltage $V_g$, ribbon width $W$, and mass-equivalent metal thickness $t$. For the thinnest UTMF under consideration ($t$=3 nm) we detect tunable plasmons in the near-IR ($\lambda$=1.5 µm) and in the mid-IR ($\lambda$=5 µm) for the narrowest ($W$=150 nm) and widest ($W$=800 nm) ribbons, respectively. The transmission and wavelength tuning ranges progressively decrease as the UTMF becomes thicker. The average transmission modulation ($\Delta T$) is 14.4%, 3.0%, and <1% for $t$=3 nm, 6 nm, and 15 nm, respectively. An analogous trend is found in the average wavelength tuning range ($\Delta\lambda/\lambda_0$), which is 13.6%, 2.1%, and <1% for $t$=3 nm, 6 nm, and 15 nm, respectively.

Our results demonstrate plasmons in ultrathin gold films, extending the plasmonic regime to infrared wavelengths and providing a broad dynamically tunable optical response. These characteristics, combined with a large-scale fabrication approach, can find applications in transparent conductors, plasmon-enhanced spectroscopy, optical biosensing, and electrochromic devices.

# FIGURES

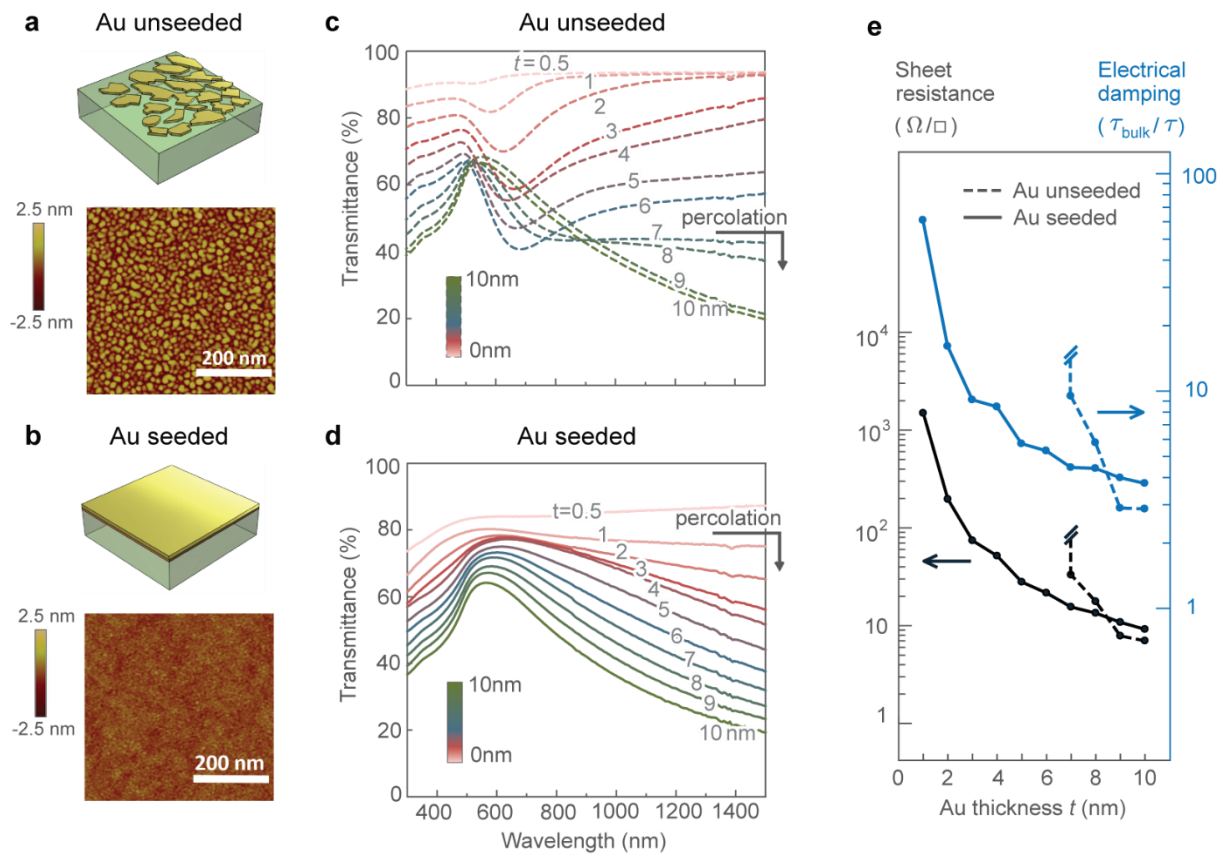

**Fig. 1. Ultrathin metal films. a, b,** Conceptual view and AFM images of gold UTMFs of mass-equivalent thickness $t \approx 3$ nm grown on 285-nm-thick native silica oxide on Silicon substrates, without and with a copper seed layer. Unseeded gold (a) grows with discontinuous island-like morphology, whereas seeded gold (b) produces continuous and relatively smooth films. **c,d,** Near-infrared and visible transmission spectrum of unseeded and seeded UTMFs grown on fused silica for different gold thicknesses $t$. The spectra of unseeded gold UTMFs (**c**), show transmission dips at around 600 nm for $t$ below 7 nm, corresponding to localized plasmon resonances of isolated gold islands and revealing an optically disconnected layer. In contrast, the spectra of seeded gold UTMFs (**d**), show progressively decreasing transmission at near-infrared wavelengths for $t$ above 1 nm, demonstrating long-range connectivity. **e,** Electrical sheet resistance and damping of seeded and unseeded gold UTMFs as a function of film thickness $t$. Unseeded gold UTMFs are conductive only for $t \geq 7$ nm (percolation thickness), whereas seeded UTMFs are conductive even down to $t=1$ nm.

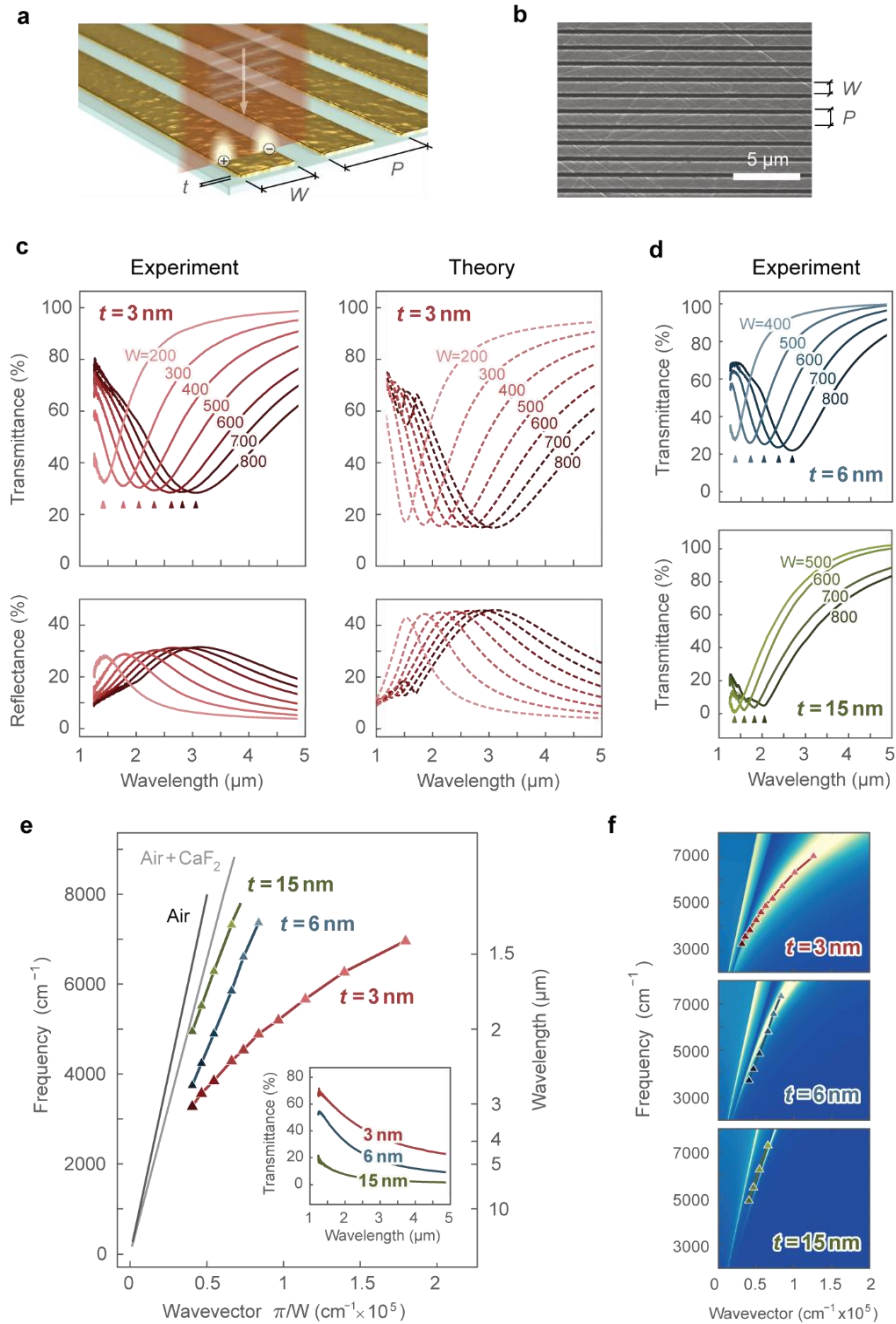

**Fig. 2. Infrared plasmons in ultrathin metal films. a,b,** Conceptual view (a) and SEM images (b) of UTMF seeded gold nanoribbons of width $W$, period $P$, and mass-equivalent thickness $t$, fabricated on CaF$_2$ substrates. **c,** Measured and simulated spectra of selected UTMF nanoribbons showing plasmonic resonances from near- to mid-infrared for different ribbon widths $W$ and fixed thickness $t$=3 nm. We use a gold Drude model with damping $\Gamma$=0.19 eV in the presented simulations (dashed curves). **d,** Measured transmission spectra of UTMF nanoribbons for $t$=6 nm and $t$=15 nm. **e,** Experimental dispersion curves of UTMF plasmons for different thicknesses obtained from the ribbon width and resonance wavelength. The thickest ribbons ($t$=15nm) approach a light-line dispersion (solid grey lines), while the thinnest ribbons ($t$=3nm) show a dispersion bending characteristic of plasmonic modes. Inset, measured transmittance spectra of continuous seeded gold UTMFs for different thicknesses. **f,** Comparison between experimental dispersion curves of UTMF plasmons and the theoretical loss function for different thicknesses. The gold Drude damping factors used in the calculations presented in the color plots are $\Gamma$=0.19 eV, 0.10 eV, and 0.07 eV for t=3 nm, 6 nm, and 15 nm, respectively, and similar results can be obtained

with larger dampings.. The experimental dispersion curve is plotted using an effective wave vector $\pi/1.25W$ (i.e., using a renormalized ribbon width $1.25W$) for $t=3$ nm.

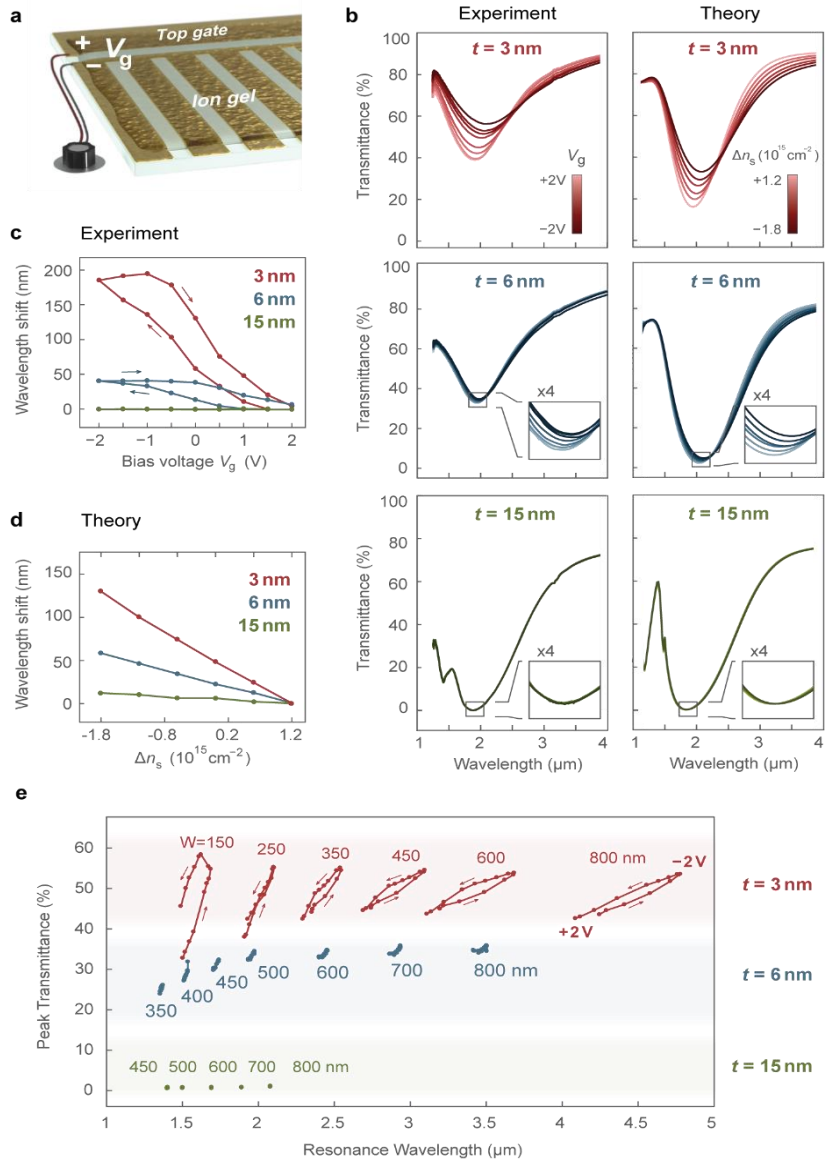

**Fig. 3. Tunable plasmons in ultrathin metal films. a,** Conceptual view of the dynamic tuning of UTMF nanoribbons by ion-gel gating. **b,** Measured and simulated transmission spectra of tunable nanoribbons resonating at $\lambda=2$ μm for thicknesses $t=3$ nm, 6 nm, and 15 nm and widths $W=250$ nm, 500 nm, and 700 nm, respectively. The experimental spectra correspond in all cases to gating voltages $V_g$ from +2 V (light color) to −2.0 V (dark color). The theoretical spectra correspond to nanoribbons with an effective surface charge density variation $\Delta n_s$ between $\Delta n_s=+1.2\times10^{15}$ carriers/cm² (light color) and $\Delta n_s=-1.8\times10^{15}$ carriers/cm² (dark color). The effective surface carrier density change, used as a fitting parameter, does not necessarily correspond to a change of bulk carrier density (number of electrons per unit volume) induced in the metal. The optical damping factors Γ for gold used in the simulations vary from 0.22 eV to 0.36 eV for $t=3$ nm, from 0.13 to 0.18 eV for $t=6$ nm, and fixed at 0.09 eV for $t=15$ nm. **c,** Measured shift of the plasmonic resonance wavelength over a full voltage cycle for different gold seeded UTMF thicknesses $t$. **d,** Simulated shift of the plasmonic resonance wavelength by varying the effective surface carrier density $\Delta n_s$. Note that the wavelength shift is referred to $\Delta n_s=+1.2\times10^{15}$ carriers/cm². **e,** Measured resonance wavelength and transmission amplitude of the plasmonic peak for different thicknesses $t$, ribbon widths $W$, and gating voltages $V_g$ in the +2 V to −2 V range.

**Acknowledgments:** We acknowledge Kavitha Kalavoor, Miriam Marchena and Johann Osmond for their help in the experiments and fruitful discussions. We acknowledge financial support from the Spanish Ministry of Economy and Competitiveness through the "Severo Ochoa" programme for Centers of Excellence in R&D (SEV-2015-0522), OPTO-SCREEN (TEC2016-75080-R), and Grant No. MAT2017-88492-R, from FundacióPrivadaCellex, from Generalitat de Catalunya through the CERCA program, from AGAUR 2017 SGR 1634, and from the European Union Seventh Framework Programme under grant agreement no. 609416 ICFONest. J. C.-F. also thanks MINECO for his research grant funded by means of the program Juan de la Cierva (Grant No. FPDI-2013-18078). F.J.G.A. acknowledges support from the European Research Council (Advanced Grant No. 789104-eNANO).

**Author contributions:** F.J.G.A. and V.P. proposed the research project. V.P. coordinated the experiments and with the help of R.A.M and D.R designed them. R.A.M and D.R with the help of J.C.-F, D.S.G, R.Yo, D.E.B and A.R carried out the experiments and characterizations. R.Yu developed the theoretical model and performed all the simulations under supervision of F.J.G.A. D.R, V.P, F.J.G.A, R.Y and R.A.M wrote the manuscript. All authors contributed to the interpretation of the results and manuscript writing.